\documentclass[11pt]{article}
\usepackage{latexsym}
\usepackage{amsmath}
\usepackage{amssymb}
\usepackage{amsthm}
\usepackage{epsfig}
\usepackage{natbib}
\usepackage{doc}  
\usepackage{url}
\usepackage[dvipsnames]{xcolor}
\usepackage{hyperref}
\usepackage{fontawesome5}

\usepackage[top=0.8in,left=0.8in,bottom=0.9in,right=0.8in]{geometry}

\usepackage{siunitx}
\usepackage{multirow}
\hypersetup{
    colorlinks=true,
    linkcolor=BrickRed,
    citecolor=BrickRed,
    urlcolor=BrickRed
}

\usepackage{float}
\usepackage{booktabs} 
\usepackage{slnotation}
\usepackage{enumitem}
\usepackage{tcolorbox}
\usepackage[toc,page]{appendix}

\newtcolorbox{mytbox}{
  colback=gray!10,      
  colframe=black,       
  boxrule=1pt,          
  arc=1mm,              
  left=6pt,right=6pt,top=6pt,bottom=6pt,
  fontupper=\footnotesize
}

\newcommand{\smfoot}{\fontsize{9.5}{11}\selectfont}

\title{Think Before You Code : A Benchmark for Testing \\ LLMs on Software Engineering Deliberation}

\author{%
  Ashman Mehra\\
  School of Computer Science\\
  Carnegie Mellon University\\
  \texttt{ashmanm@andrew.cmu.edu}
}

\date{} 

\begin{document}

\maketitle

\begin{abstract}
\noindent Think Before You Code (TBYC) is a benchmark pipeline for studying whether code-focused language models can reason about software issues in ways that resemble human engineering deliberation. The benchmark curates GitHub issue discussions into structured records, extracts human discussion artifacts such as problems, solutions, constraints, rationales, and trade-offs, and then prompts models to generate issue-level “thoughts” using retrieved code context and few-shot examples. It converts model responses back into the same artifact space and evaluates them against human annotations using complementary metrics for artifact type detection, metadata matching, tag overlap, and semantic summary similarity. Beyond scoring, this project includes visualization, leaderboard, and baseline tools that make it possible to compare models, prompting strategies, and extraction methods across repositories and issues. Think Before You Code is an end-to-end evaluation environment for measuring how closely LLM software reasoning aligns with human strategic judgment.
\end{abstract}

\textbf{\texttt{CODE FOR THE PROJECT CAN BE FOUND HERE -} \href{https://github.com/ashman17/ThinkBeforeYouCode}{\textcolor{black}{\faGithub}\ \texttt{ashman17/ThinkBeforeYouCode}}}

\section{Introduction}
\vspace{-5mm}
This project addresses the use case of evaluating whether code LLMs can do more than generating plausible code and instead reason through software engineering problems the way human developers do. In real development, strong performance depends on understanding bug reports, surfacing constraints, weighing trade-offs, asking clarifying questions, using code context, and proposing grounded next steps. Think Before You Code (TBYC) turns that broader reasoning process into something measurable by building a pipeline from GitHub issue discussions to structured human reasoning artifacts, then comparing model behavior against those artifacts. The intended downstream users are researchers and practitioners who want to benchmark coding models not just on final answers, but on the quality of their intermediate engineering judgment.

The main research questions are (modified from the proposal): 
\begin{enumerate}[nosep]
    \item Can code LLMs reconstruct the kinds of reasoning humans express in issue discussions?
    \item How much does retrieved code context help compared with no-context, BM25-only, or few-shot baselines?
    \item Which parts of human reasoning are easiest or hardest for models to match, such as identifying artifact types, preserving metadata, assigning tags, or producing semantically aligned summaries?
\end{enumerate}

This research project studies these questions by extracting human artifacts from discussions, generating model “issue thoughts,” re-extracting comparable artifacts from model outputs, and scoring agreement with a suite of metrics and visualizations.

Users interact with the application primarily through the CLI pipeline. They can fetch and normalize GitHub issues, extract human discussion artifacts, retrieve relevant code chunks, run different reasoning-generation settings, derive comparable artifacts from model outputs, compute metrics, and generate graphs, leaderboards, and an HTML viewer. In that sense, the application behaves like an end-to-end research workbench rather than a consumer-facing product: the user supplies repositories and models, and the system produces structured benchmark outputs for analysis and comparison.

The project engages with course material in several direct ways. It connects to code-writing assistants by focusing on software reasoning rather than raw code completion, and it uses the science of prompting through structured prompts for both artifact extraction and issue-thought generation. It draws on learning representations and embeddings, retrieval, and retrieval-augmented generation through CodeBERT embeddings, BM25, FAISS indexing, and hybrid code-context retrieval. It also engages LLMs for evaluation by using model outputs as analyzable artifacts and scoring them against human annotations, while raising questions related to deployment and harms - a coding assistant that sounds confident but misses human constraints or trade-offs can still be unreliable in practice.

\section{Literature Review}
\vspace{-5mm}
There is not yet a benchmark that directly evaluates whether LLMs can reason through real GitHub issues in terms of design choices, feasibility, prioritization, and trade-offs, but several related efforts study code generation, repository-level reasoning, and software engineering performance.

\subsection{Code Generation}
SWE-Bench (\cite{jimenez2024swebench}) evaluates models on real GitHub issues paired with gold patches and test suites, measuring whether models can resolve issues through execution-grounded patch generation. It is highly realistic, but focuses on patch correctness rather than the quality of engineering deliberation. LiveCodeBench (\cite{ICLR2025_94074dd5}) provides a contamination-resistant and continuously updated benchmark for code generation, self-repair, and execution on recent programming tasks. Its emphasis is on robustness and iterative problem solving, not human-like reasoning about software decisions. HumanEval and Codex (\cite{chen2021evaluating}) established execution-based code evaluation through hidden unit tests and introduced \texttt{pass@k} as a standard metric for code synthesis. The benchmark measures whether generated code works, but not whether the reasoning behind it reflects human engineering judgment. The survey in \cite{jiang2024survey} reviews major techniques in code LLMs, including instruction tuning, retrieval augmentation, tool use, and RLHF, while also summarizing common evaluation strategies. It highlights that most code benchmarks prioritize functional correctness over higher-level reasoning such as prioritization, feasibility, and trade-off analysis.

\subsection{Code Reasoning}
RepoCoder (\cite{zhang-etal-2023-repocoder}) introduces iterative retrieval-augmented generation for repository-level code completion and shows that cross-file context improves performance. Its evaluation centers on completion quality rather than reasoning over issue discussions. CrossCodeEval (\cite{ding2023crosscodeeval}) studies cross-file code completion across multiple programming languages, requiring models to reason over inter-file dependencies. It captures repository structure well, but still evaluates completion accuracy rather than deliberative engineering behavior. CRUXEval (\cite{gu2024cruxeval}) focuses on reasoning about existing code by testing models on execution behavior and program semantics. This is closer to code understanding, but it does not evaluate reasoning about collaborative software discussions. CodeT (\cite{chen2022codet}) improves code generation by jointly generating code and tests, then selecting candidates through verification. It shows the value of reasoning with validation, though the target remains final solution correctness. CodeXGLUE (\cite{lu2021codexglue}) is a broad benchmark suite covering code understanding and generation tasks such as completion, translation, summarization, and defect detection. It demonstrates the breadth of code intelligence tasks, but not issue-level engineering deliberation. SWE-agent (\cite{xia2024sweagent}) studies agentic systems that interact with repositories and tools to resolve GitHub issues. It moves closer to realistic software workflows, but still evaluates end-to-end task completion more than human-like intermediate reasoning.

\subsection{Industry Efforts}
OpenAI Codex (\cite{openai2021codex}) helped popularize natural-language-to-code generation and demonstrated the practical value of large code models. Its framing is centered on synthesis capability and productivity rather than explicit benchmarking of engineering judgment. GitHub Copilot (\cite{githubcopilotdocs}) brings code completion and chat-based coding assistance into everyday developer workflows. It is an important practical system, but public evaluation focuses more on usefulness and productivity than on reasoning quality. Devin (\cite{cognition2024devin}) represents the shift toward autonomous software agents that can plan, code, debug, and iterate across tools. However, public descriptions emphasize task completion and autonomy claims rather than alignment with human software reasoning.

\subsection{Research Gap}
Across these works, the dominant evaluation targets are code correctness, repository-aware completion, and end-to-end software task performance. What remains underexplored is whether models can reason like human engineers before implementation by surfacing trade-offs, feasibility limits, blockers, uncertainty, and prioritization. This is an important capability because LLMs should not only be able to ``reproduce" the known solutions but be able to ``produce" these solutions. This will help LLMs solve unsolved problems.

The benchmark addresses this gap by shifting evaluation from \emph{whether code works} to \emph{whether the model’s reasoning aligns with human engineering deliberation} in real GitHub issue discussions. This makes it complementary to existing code-generation and software-engineering benchmarks rather than a replacement for them.

\section{Method}
\vspace{-5mm}
The method evaluates whether code-oriented LLMs can reason about real software engineering issues in ways that align with human discussion, rather than only producing correct code. The full pipeline is organized into six  (Figure 1): (1) issue and repository collection, (2) gold artifact extraction from human discussions, (3) retrieval-augmented context construction, (4) LLM inference for issue-level reasoning, (5) artifact extraction from LLM outputs, and (6) evaluation of alignment between gold and model-generated artifacts. The overall design converts unstructured issue discussions into a structured representation of engineering reasoning, then measures how closely LLM-generated reasoning matches that representation.

\begin{figure}[H]
    \centering
    \includegraphics[width=0.8\linewidth]{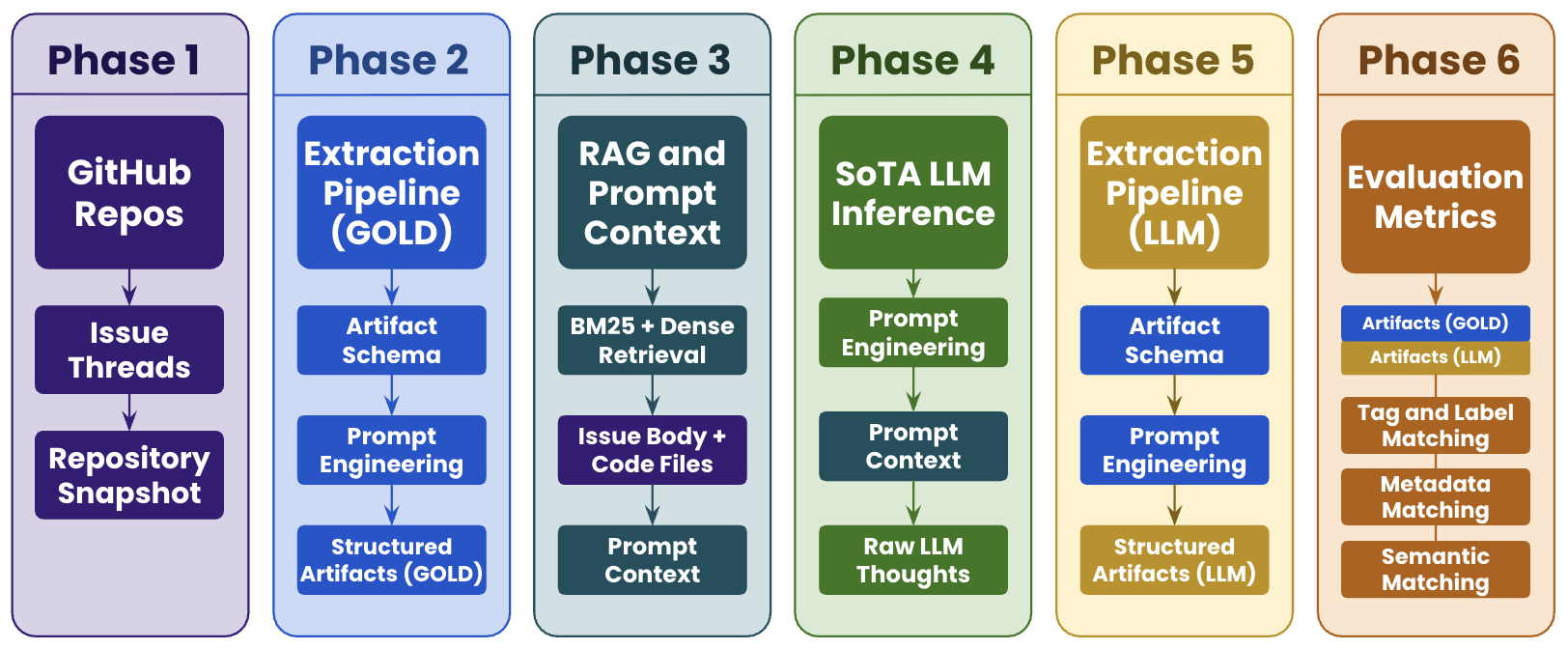}
    \caption{Phase-by-phase breakdown of the Think Before You Code benchmark.}
    \label{fig:placeholder}
\end{figure}

\subsection{Phase 1: Repository and Issue Curation}
The pipeline begins by selecting public GitHub repositories with active maintainers, reputational significance, and sufficiently rich issue discussions. To maximize the value of human deliberation signal, issues with dense commentary, diversity in issue types, and closed status are prioritized so that each discussion is complete. Recent issues are also emphasized to reduce the risk of benchmark contamination from model pretraining data.

For each repository, the GitHub GraphQL API is used to collect issue-centered snapshots. Each snapshot includes the issue title and body, comment thread, labels, state, author metadata, and timeline events. These raw payloads are stored for auditability and future reprocessing. The raw snapshots are then normalized into curated issue records containing the input vector (title and body), the deliberation thread, file references mentioned in the discussion, taxonomic metadata, timeline events, and actor typology. The current curated dataset in the repository contains seven repositories and \(\approx\)1000 issues in total as shown in Table 1. This construction yields a multi-repository benchmark over real issue threads rather than synthetic problem statements.

\begin{table}[h]
\label{tab:repo}
    \centering
    \caption{Target repositories along with number of issues parsed for each for curating the dataset.}
    \vspace{2mm}
    \small
    \begin{tabular}{llcp{7cm}}
    \hline
    \hline
    \textbf{Repository} & \textbf{Category} & \textbf{Issues} & \textbf{Why It’s Important} \\
    \hline
    \hline
    Bitcoin & Distributed Systems & 120 & Deep discussions on consensus, security, decentralization, and protocol-level trade-offs often with rigorous reasoning and long-term design implications \\
    \hline
    Kubernetes & Systems & 81 & Rich discussions on scalability, reliability, and distributed system trade-offs \\
    \hline
    Rust & Prog. Language & 92 & Extremely structured design debates with clear alternatives and rationales \\
    \hline
    PyTorch & ML Framework & 76 & Frequent discussions on performance vs usability and API evolution \\
    \hline
    Bazel & Developer Tools & 97 & Deep reasoning about dependency management and build performance \\
    \hline
    FastAPI & Web Framework & 57 & Clean, well-structured discussions led by maintainers—great for annotation \\
    \hline
    Hugging Face & ML / LLM & 43 & Rapid iteration with strong focus on API design and usability trade-offs \\
    \hline
    PostgreSQL & Database & 75 & High-quality discussions on correctness, indexing, and query optimization \\
    \hline
    React & UI Framework & 129 & Focus on API ergonomics and backward compatibility decisions \\
    \hline
    Redis & Database & 89 & Clear trade-offs between latency, memory usage, and data structures \\
    \hline
    \hline
    \end{tabular}
    \normalsize
    \label{tab:target-repos}
\end{table}

\subsection{Phase 2: Gold Artifact Extraction from Human Discussion}
The second phase converts human issue discussions into structured reasoning artifacts. The goal is to encode the latent engineering content of a discussion in a machine-comparable schema. An artifact schema is defined with four required fields: \texttt{type}, \texttt{summary}, \texttt{tags}, and \texttt{metadata}. The current implementation supports 31 artifact types, including \texttt{problem\_statement}, \texttt{proposed\_solution}, \texttt{trade\_off\_argument}, \texttt{rationale}, \texttt{constraint}, \texttt{question}, \texttt{task\_assignment}, \texttt{status\_update}, \texttt{blocking\_issue}, and \texttt{dependency}. Each type has a fixed metadata schema, such as \texttt{problem\_kind}, \texttt{affected\_area}, and \texttt{user\_or\_system\_impact} for problem statements, or \texttt{options\_compared}, \texttt{comparison\_axes}, and \texttt{preferred\_option} for trade-off arguments. Refer to the Appendix (Table 6) for complete list with descriptions.

Tags provide finer-grained labels that capture bug classes, platform details, infrastructure signals, and software domains. The current schema contains 81 allowed tags (Appendix-B for complete list), including labels such as \texttt{regression}, \texttt{flaky\_test}, \texttt{runtime\_error}, \texttt{performance\_issue}, \texttt{compatibility\_issue}, \texttt{linux}, \texttt{macos}, \texttt{docker}, \texttt{frontend}, \texttt{backend}, \texttt{rpc}, and \texttt{fix\_proposed}. Each artifact also includes a short summary that must state the subject, event, and key technical detail from the underlying comment.

To extract these artifacts from human discussion, a structured extraction pipeline is run over individual comments. The extraction prompt (Appendix-A) instructs the model to return strict JSON only, choose exactly one of the allowed artifact types, provide one or more supported tags from the controlled vocabulary, and fill in the fixed metadata keys for that type without null values or extra fields. The prompt is added to the Appendix for reference. The implementation uses Pydantic schemas together with structured decoding through Outlines/Ollama to enforce well-formed JSON output. A signifiacnt limitation of this method is that the artifacts can have errors due to the LLM annotation design. There is no automatic method that can eliminate the need of an LLM. A smaller human verified dataset will be an ideal solution and is part of the future works. For now, to combat the problem of LLM bias, artifacts from multiple open-source models (Gemma3, Gemma4, Llama) are combined to increase diversity and recall while reducing bias. These open-source models do not compete in the the final experiments and leaderboard.

\begin{figure}
    \centering
    \includegraphics[width=0.8\linewidth]{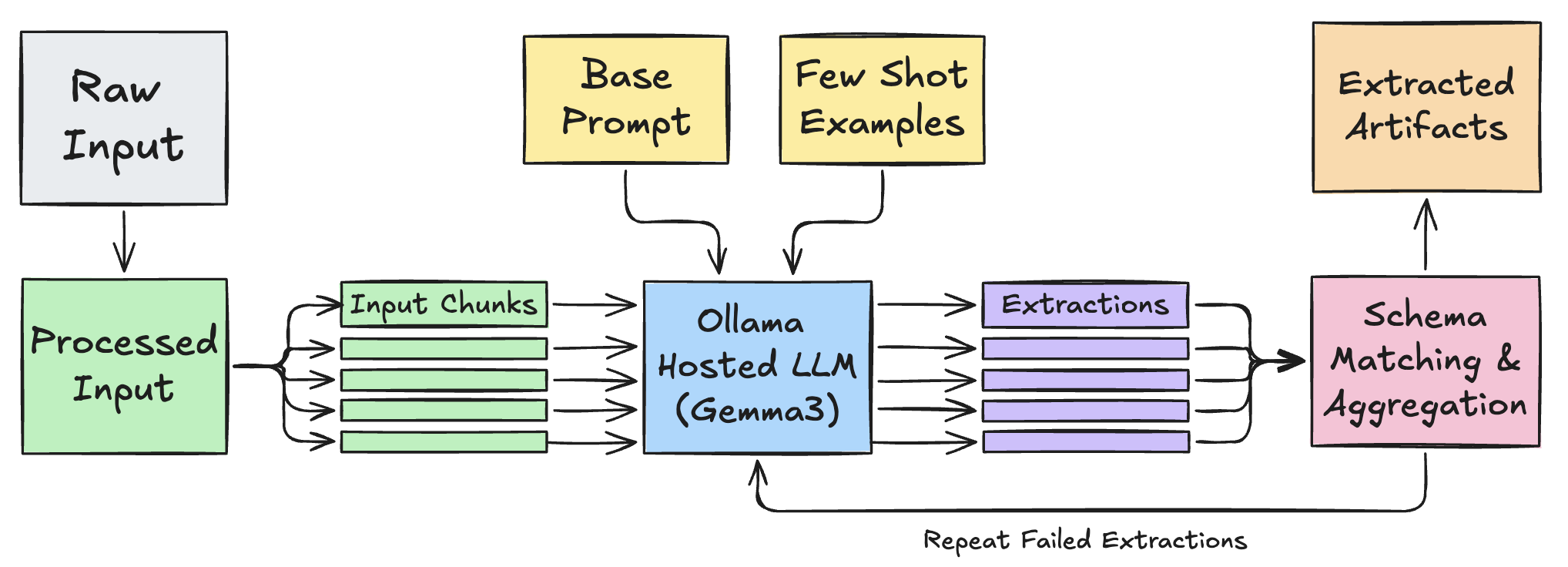}
    \caption{Overview of artifact extraction pipeline. The input is batched into chunks and passed through multiple open-source models (Gemma and Llama family) to produce the extractions which undergo a schema matching to yield the final extracted artifacts.}
    \label{fig:placeholder}
\end{figure}

\subsection{Phase 3: Retrieval-Augmented Context Construction}
The third phase builds repository context for issue-conditioned reasoning. The core question in this phase is what information a model should see when asked to reason like a human engineer. The benchmark intentionally provides the issue body and retrieved code context, but not the original human comments, since the model’s task is to independently reconstruct human-like deliberation rather than paraphrase the discussion.

{\sloppy
To supply code context, the pipeline first resolves the repository snapshot corresponding to the latest default-branch commit before the earliest issue timestamp in the curated set. In the repository snapshot, non-source files are pruned and source files are split into code chunks, ideally at the symbol level and otherwise in fixed-size windows. Each issue query is formed from the issue title and body.}

Retrieval uses a hybrid sparse--dense architecture (ablation in results). For sparse retrieval, each chunk is indexed with BM25. Given issue query $q$ and chunk $d$, the Okapi BM25 score is:
\[
s_{\text{bm25}}(d,q)
= \sum_{t \in q}
\operatorname{IDF}(t)
\cdot
\frac{f(t,d)(k_1+1)}{f(t,d)+k_1\left(1-b+b\frac{|d|}{\operatorname{avgdl}}\right)}.
\]
This component is especially useful for exact matches on identifiers, file names, APIs, and issue-specific terminology. For dense retrieval, both queries and code chunks are encoded with a CodeBERT-family encoder. After mean pooling and L2 normalization, chunk relevance is scored by cosine similarity:
\[
s_{\text{dense}}(d,q)
= \langle \hat{\mathbf{e}}(d), \hat{\mathbf{e}}(q) \rangle
= \cos\big(\hat{\mathbf{e}}(d),\hat{\mathbf{e}}(q)\big).
\]
This branch helps when lexical overlap is weak but semantic intent is similar. The BM25 and dense rankings are fused using Reciprocal Rank Fusion (RRF):
\[
\operatorname{RRF}(d)
= \sum_{r \in \{\text{bm25},\text{dense}\}}
\frac{1}{k + \operatorname{rank}_r(d)}.
\]
The implementation uses $k=60$ by default. This hybrid strategy combines lexical precision, semantic generalization, and rank-level robustness. The top retrieved chunks are then formatted as prompt context in the form \texttt{[file:start-end (symbol)]} followed by chunk text.

\subsection{Phase 4: LLM Inference for Issue Thoughts}
In the fourth phase, modern LLMs are prompted to generate issue-level reasoning traces. The prompt asks the model to ``read the issue and think like a human engineer,'' then produce at least 20 short comments, one per line, each formatted as \texttt{[Topic] comment} (full prompt in Appendix-A). The topic inventory is fixed and mirrors the artifact schema categories, including Problem Statement, Proposed Solution, Trade-off Argument, Rationale, Constraint, Question, Suggestion, Task Assignment, Status Update, and Dependency. Conflicting or partially unrelated comments are explicitly allowed so that the model can surface multiple lines of thought rather than converge on a single polished answer.

The prompt consists of the issue title and body plus the retrieved code context from Phase 3. In additional baseline settings, the system can omit code context entirely or include few-shot examples derived from repository-matched extractions. Thus, the repository supports three main inference settings: the original RAG condition, a few-shot condition, and a no-context condition.

The evaluation artifacts cover nine models: \texttt{gemini-2.5-flash}, \texttt{gemini-3.1-flash}, \texttt{gpt-5.4}, \texttt{gpt-5}, \texttt{gpt-4.1-mini}, \texttt{claude-opus-4.6}, \texttt{claude-sonnet-4.6}, \texttt{qwen2.5:7b-instruct}, \texttt{deepseek-r1:8b}. This mix includes both open-weight and API-based frontier models, allowing comparison across provider families and capability tiers.

\subsection{Phase 5: Artifact Extraction from LLM Responses}
The fifth phase maps raw LLM thoughts back into the same structured space used for gold artifacts. Each model response is split into comment-like units, and the artifact extraction system is re-run over those units using the same artifact types, tag vocabulary, metadata fields, and structured decoding rules as in the gold pipeline. This makes the representation on the human side and the model side directly comparable.

This symmetry is important. Instead of comparing free-form model outputs to free-form human discussion, structured artifacts are compared to structured artifacts. The result is a two-stage benchmark: first generate issue reasoning, then evaluate the extracted reasoning content. This design separates the quality of deliberation from the surface style of generation.

\subsection{Dataset}
The benchmark uses real GitHub issue discussions as its primary data source. The main dataset curation criteria are:
\begin{itemize}[nosep]
\item issues with dense commentary, so there is enough deliberative signal to recover,
\item diversity in issue types, to avoid overfitting to a narrow class of bugs or requests,
\item well-maintained and reputable repositories, so the discussions reflect realistic engineering practice,
\item closed issues, so the discussion thread is sufficiently complete,
\item recent issues, to reduce the likelihood of leakage from model pretraining corpora.
\end{itemize}

In addition to the processed issue records, the repository includes raw issue snapshots, code snapshot, extraction outputs, retrieval indexes, model responses, derived artifact files, metrics, baseline comparisons, and visualization manifests. This makes the benchmark reproducible end-to-end.

\begin{table}[H]
\centering
\caption{Notations for Metrics and Evaluation}
\vspace{1mm}
\small
\setlength{\tabcolsep}{0pt}
\renewcommand{\arraystretch}{1.2}
\begin{tabular}{p{0.35\linewidth} p{0.6\linewidth}}
\hline
\hline
\textbf{Symbol} & \textbf{Description} \\
\hline
\hline
$H_i$ & Set of human artifact types present in issue $i$ \\
\hline
$L_i$ & Set of LLM artifact types present in issue $i$ \\
\hline
$M_i = H_i \cap L_i$ & Set of matching artifact types in issue $i$ \\
\hline
$F_{i,t}^{H},\; F_{i,t}^{L}$ & Human and LLM metadata field-name sets for issue $i$, type $t$ \\
\hline
$V_{i,t,f}^{H} = (v_1,\dots,v_m)$ & Human metadata values for issue $i$, type $t$, field $f$ \\
\hline
$V_{i,t,f}^{L} = (u_1,\dots,u_n)$ & LLM metadata values for issue $i$, type $t$, field $f$ \\
\hline
$G_{i,t}^{H},\; G_{i,t}^{L}$ & Human and LLM tag sets for issue $i$, type $t$ \\
\hline
$S_{i,t}^{H} = (s_1,\dots,s_a)$ & Human summary strings for issue $i$, type $t$ \\
\hline
$S_{i,t}^{L} = (r_1,\dots,r_b)$ & LLM summary strings for issue $i$, type $t$ \\
\hline
$\operatorname{combine}(x_1,\dots,x_k)$ & Newline-joined concatenation of non-empty summary strings \\
\hline
$\operatorname{ratio}(x,y)$ & \(x/y\) if \(y>0\), 1 if \(y=x=0\), and \(0\) otherwise \\
\hline
$\operatorname{harm}(p,r)$ & \(0\) if \(p+r = 0\), \(\frac{2pr}{p+r}\) otherwise \\
\hline
\end{tabular}
\end{table}

\subsection{Evaluation Scheme}
The evaluation compares gold human artifacts and LLM-generated artifacts along four complementary dimensions: type matching, metadata matching, tag matching, and semantic summary matching. Together, these metrics assess not only whether the model identified the right reasoning categories, but also whether it preserved the fine-grained content and meaning of human discussion.

\subsubsection{Type Matching}
Type matching measures whether the model predicts the correct artifact types for each issue. Let $H_i$ be the set of human artifact types for issue $i$ and $L_i$ the set of LLM artifact types. Then:
\[
\mathrm{TP}_i = |H_i \cap L_i|,\quad
\mathrm{FP}_i = |L_i \setminus H_i|,\quad
\mathrm{FN}_i = |H_i \setminus L_i|.
\]
\[
\mathrm{Precision}_i = \operatorname{ratio}(\mathrm{TP}_i, |L_i|),\quad
\mathrm{Recall}_i = \operatorname{ratio}(\mathrm{TP}_i, |H_i|),
\]
\[
\mathrm{F1}_i = \operatorname{harm}(\mathrm{Precision}_i,\mathrm{Recall}_i),\quad
\mathrm{Jaccard}_i = \operatorname{ratio}(\mathrm{TP}_i, |H_i \cup L_i|).
\]
These scores measure whether the model identifies the same kinds of reasoning present in the human discussion.

\subsubsection{Tag Matching}
Tag matching evaluates whether the detailed labels attached to correctly matched types agree. For issue $i$ and matched type $t$, let $G_{i,t}^{H}$ and $G_{i,t}^{L}$ denote the human and LLM tag sets. Then:
\[
\mathrm{Precision}_{i,t}=\operatorname{ratio}(|G_{i,t}^{H}\cap G_{i,t}^{L}|,|G_{i,t}^{L}|) \quad\quad \mathrm{Recall}_{i,t}=\operatorname{ratio}(|G_{i,t}^{H}\cap G_{i,t}^{L}|,|G_{i,t}^{H}|),
\]
\[
\mathrm{F1}_{i,t}=\operatorname{harm}(\mathrm{Precision}_{i,t},\mathrm{Recall}_{i,t}),
\quad
\mathrm{Jaccard}_{i,t}=\operatorname{ratio}(|G_{i,t}^{H}\cap G_{i,t}^{L}|,|G_{i,t}^{H}\cup G_{i,t}^{L}|).
\]
This captures whether the model matches the detailed characterization of an artifact once it has identified the right artifact type.

\subsubsection{Metadata Matching}
Metadata matching compares the field values attached to each artifact type and field. Since metadata values are free-form strings after normalization, exact equality would be too brittle. Both hard and soft matching scores are therefore computed.

For a given issue $i$, type $t$, and field $f$, let $V_{i,t,f}^{H}$ and $V_{i,t,f}^{L}$ be the human and LLM metadata values. Pairwise phrase similarities are computed, one-to-one matches are formed greedily, and a hard true positive is counted when the similarity exceeds threshold $\tau$ (default $0.82$). Hard metrics are:
\[
\mathrm{Precision}_{i,t,f}=\operatorname{ratio}(\mathrm{TP}_{i,t,f},|V_{i,t,f}^{L}|),\quad
\mathrm{Recall}_{i,t,f}=\operatorname{ratio}(\mathrm{TP}_{i,t,f},|V_{i,t,f}^{H}|),
\]
\[
\mathrm{F1}_{i,t,f}=\operatorname{harm}(\mathrm{Precision}_{i,t,f},\mathrm{Recall}_{i,t,f}),
\quad
\mathrm{Jaccard}_{i,t,f}=\operatorname{ratio}(\mathrm{TP}_{i,t,f},\mathrm{TP}_{i,t,f}+\mathrm{FP}_{i,t,f}+\mathrm{FN}_{i,t,f}).
\]

Soft metrics are also computed by assigning each human value and each LLM value its best similarity match on the opposite side. If $\Sigma^L_{i,t,f}$ and $\Sigma^H_{i,t,f}$ denote the summed best-match similarities, then:
\[
\mathrm{SoftPrecision}_{i,t,f}=
\begin{cases}
1 & |V_{i,t,f}^{L}|=0\\
\Sigma^{L}_{i,t,f}/|V_{i,t,f}^{L}| & \text{otherwise}
\end{cases}
\quad\quad
\mathrm{SoftRecall}_{i,t,f}=
\begin{cases}
1 & |V_{i,t,f}^{H}|=0\\
\Sigma^{H}_{i,t,f}/|V_{i,t,f}^{H}| & \text{otherwise}
\end{cases}
\]
\[
\mathrm{SoftF1}_{i,t,f}=\operatorname{harm}(\mathrm{SoftPrecision}_{i,t,f},\mathrm{SoftRecall}_{i,t,f}).
\]
The implementation supports multiple similarity backends, including token F1, sequence ratio, token Jaccard, character 3-gram Jaccard, token containment, and a max-over-backends ensemble. Refer to Appendix-C for more details.

\subsubsection{Semantic Summary Matching}
Summary matching evaluates whether the semantic content of human and LLM artifact summaries aligns for matched types. For each issue $i$ and matched type $t$, all human summaries of that type are concatenated into $\bar{S}_{i,t}^{H}$ and all LLM summaries into $\bar{S}_{i,t}^{L}$. These aggregated summaries are scored using three complementary semantic similarity measures.

First, CodeBERT cosine similarity is computed from mean-pooled encoder embeddings:
\[
\mathrm{CodeBERTCosine}(x,y)=
\frac{\phi(x)\cdot\phi(y)}{\|\phi(x)\|_2\|\phi(y)\|_2}.
\]

Second, BERTScore is computed by contextual token alignment:
\[
\mathrm{BERTScorePrecision}(x,y)=\frac{1}{n}\sum_{j=1}^{n}\max_k A_{jk},
\quad
\mathrm{BERTScoreRecall}(x,y)=\frac{1}{m}\sum_{k=1}^{m}\max_j A_{jk},
\]
\[
\mathrm{BERTScoreF1}(x,y)=\operatorname{harm}(\mathrm{BERTScorePrecision}(x,y),\mathrm{BERTScoreRecall}(x,y)).
\]

Third, a BLEURT score $\beta(x,y)$ is computed, optionally with raw, clipped, or sigmoid-normalized post-processing. Refer to Appendix for equations and details of how BLEURT is calculated. These three measures together capture embedding-level similarity, token-level semantic coverage, and learned text quality. Refer to Appendix-D for details on BLUERT calculation

\subsubsection{Coverage-Aware Evaluation}
A key challenge is that semantic summary comparison is only defined for matched artifact types. A model might achieve strong semantic similarity on a small number of matched types while missing many human reasoning categories entirely. To account for this, issue-level coverage metrics are computed:
\[
\mathrm{CovPrecision}_i=\operatorname{ratio}(|H_i \cap L_i|,|L_i|),\quad
\mathrm{CovRecall}_i=\operatorname{ratio}(|H_i \cap L_i|,|H_i|),
\]
\[
\mathrm{CovF1}_i=\operatorname{harm}(\mathrm{CovPrecision}_i,\mathrm{CovRecall}_i).
\]
Summary quality is then penalized by coverage recall:
\(
q^{\mathrm{pen}}_i = \operatorname{PenaltyApply}(\overline{q}_i,\mathrm{CovRecall}_i).
\)

This prevents overestimating models that write high-quality summaries for only a narrow subset of the human reasoning space.

\section{Experimental Results}
\vspace{-5mm}

This section presents the empirical evaluation of the nine state-of-the-art models using the benchmark. The primary objective of these experiments is to determine whether current large language models (LLMs) can replicate the nuanced deliberation found in professional software engineering discussions. The results indicate that while models demonstrate a baseline ability to identify technical problems, they remain significantly limited in their capacity for deep engineering reasoning and structural deliberation.

\begin{figure}[H]
    \centering
    \includegraphics[width=0.9\linewidth]{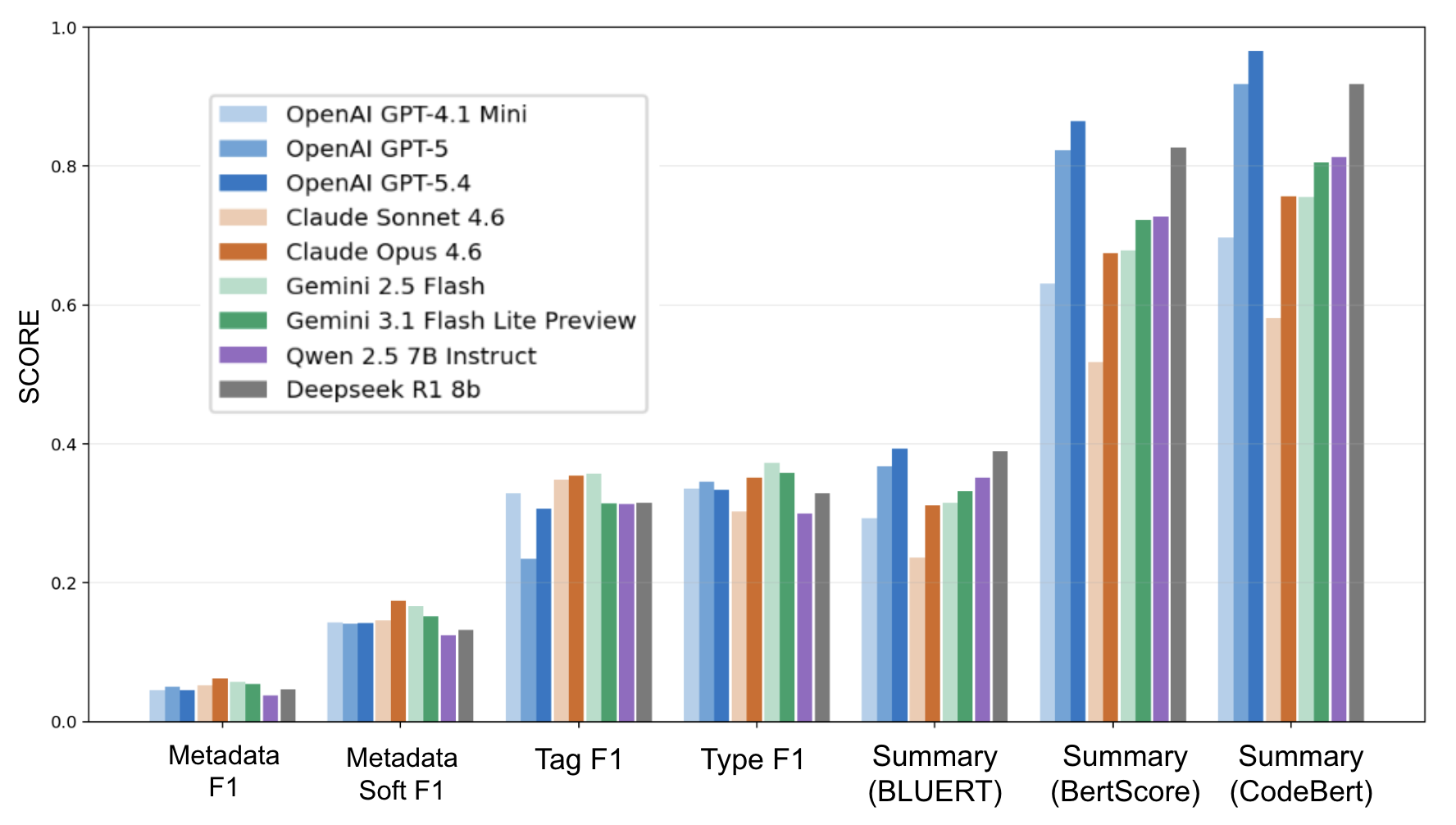}
    \caption{Scores for the 9 models across the 7 metrics. Metadata F1 scores are below 0.2 which suggests that the models do not know the correct ideas under a topic and use generic terms. Tag and Type F1 scores are also lower in general. The models show considerably higher scores on summary. This suggests that models often respond in generic terms based on their training data and are not able to narrow down to the exact topics to think about.}
    \label{fig:placeholder}
\end{figure}

\subsection{Taxonomy and Reasoning Type Discrepancies}
An analysis of per-type F1 scores (Figure 3) reveals ``design choice blindness" within current models. While models are relatively proficient at identifying \texttt{problem\_statement} and \texttt{proposed\_solution} artifacts, performance plummeted when encountering \texttt{trade\_off\_argument} and \texttt{design\_decision} categories (Figure 4). This discrepancy highlights that LLMs are currently optimized for execution-based tasks---identifying what is broken and how to fix it---rather than the deliberative reasoning required to assess feasibility or prioritize competing implementation strategies. In discussing the correct ``types'' of topics (Figure 5) , models are somewhat similar in their behavior but remain far from perfect in capturing the breadth of human deliberation.

\begin{figure}[h]
    \centering
    \includegraphics[width=0.9\linewidth]{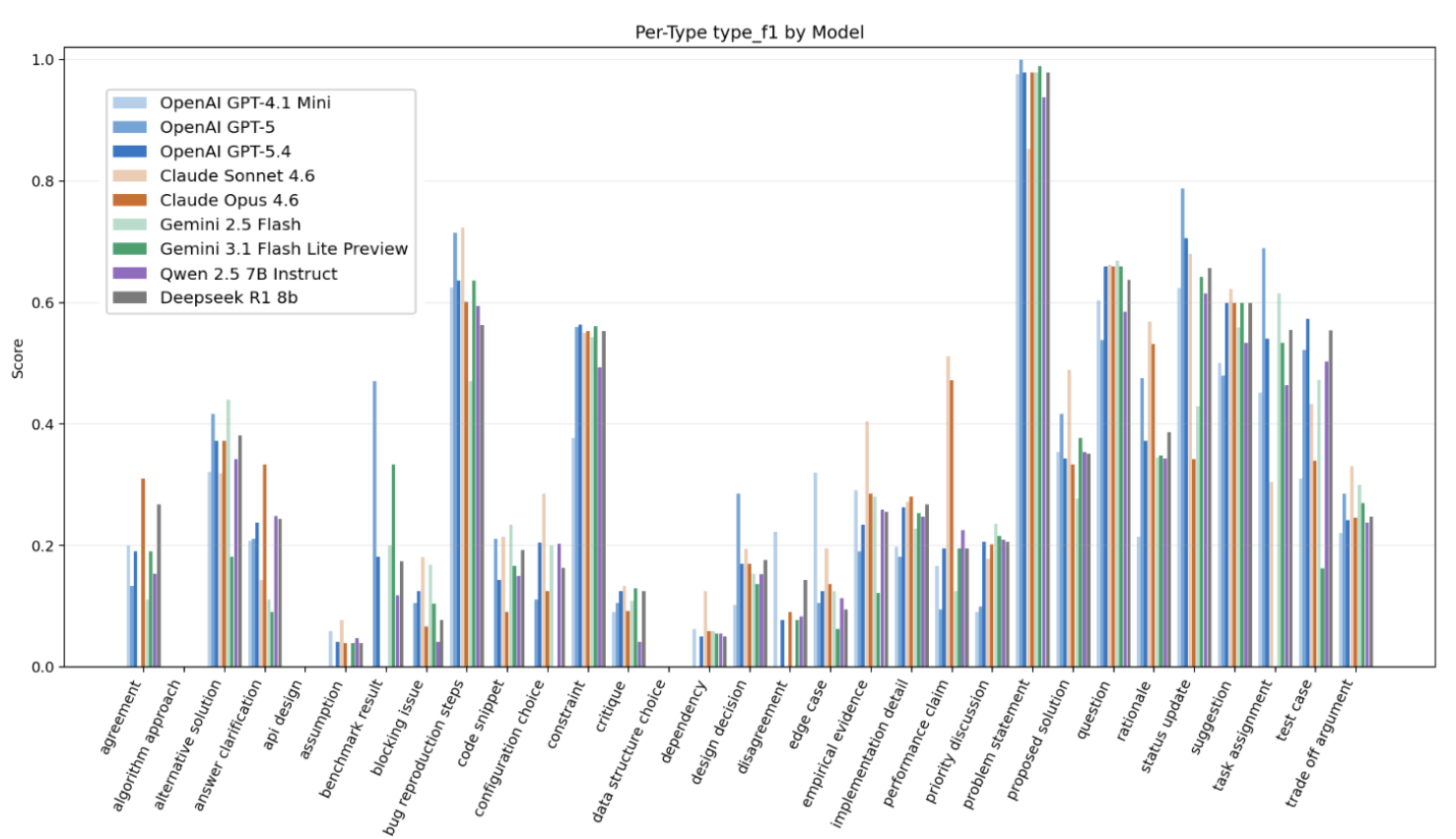}
    \caption{Figure shows the Type F1 scores. The models where asked to discuss about one of the 32 types (see Appendix for complete list). The granular results show that the state of the art models are signifiacntly poor when thinking about trade-offs, data structures, and design choices. This confirms the hypothesis that most code LLMs work well because they reproduce previously seen code and not due to a superior (and human-like) reason ability.}
    \label{fig:placeholder}
\end{figure}

\subsection{Structural Deliberation and the Metadata Gap}
A critical failure point identified in the evaluation is the extraction of structured metadata. As shown in the per-issue metadata F1 results (Figure 5), scores are universally low, often failing to cross the 0.20 threshold even for soft-matching metrics. This indicates that while a model might correctly identify a \texttt{trade\_off\_argument}, it struggles to extract the specific \texttt{options\_compared} or the \texttt{preferred\_option} with the same precision as a human developer. This structural conclusion gap represents the largest scope for future improvement, as it requires the model to not only recognize a discussion but to derive actionable, structured conclusions from it.

\subsection{Semantic Alignment and Stability}
An ensemble of semantic similarity measures was employed to evaluate the alignment of model ``thoughts" with human summaries. The findings confirm that BLEURT serves as a significantly harder metric than BERTScore or CodeBERT. The former's learned sensitivity to text quality and technical logic frequently penalized models that, according to token-alignment metrics like BERTScore, appeared to be performing well. Furthermore, newer iterations of OpenAI and DeepSeek models demonstrated higher stability (Figure 5) , characterized by lower standard deviation in summary metrics across all repositories, suggesting a maturing capability in consistent text generation.

\begin{figure}[h]
    \centering
    \includegraphics[width=0.9\linewidth]{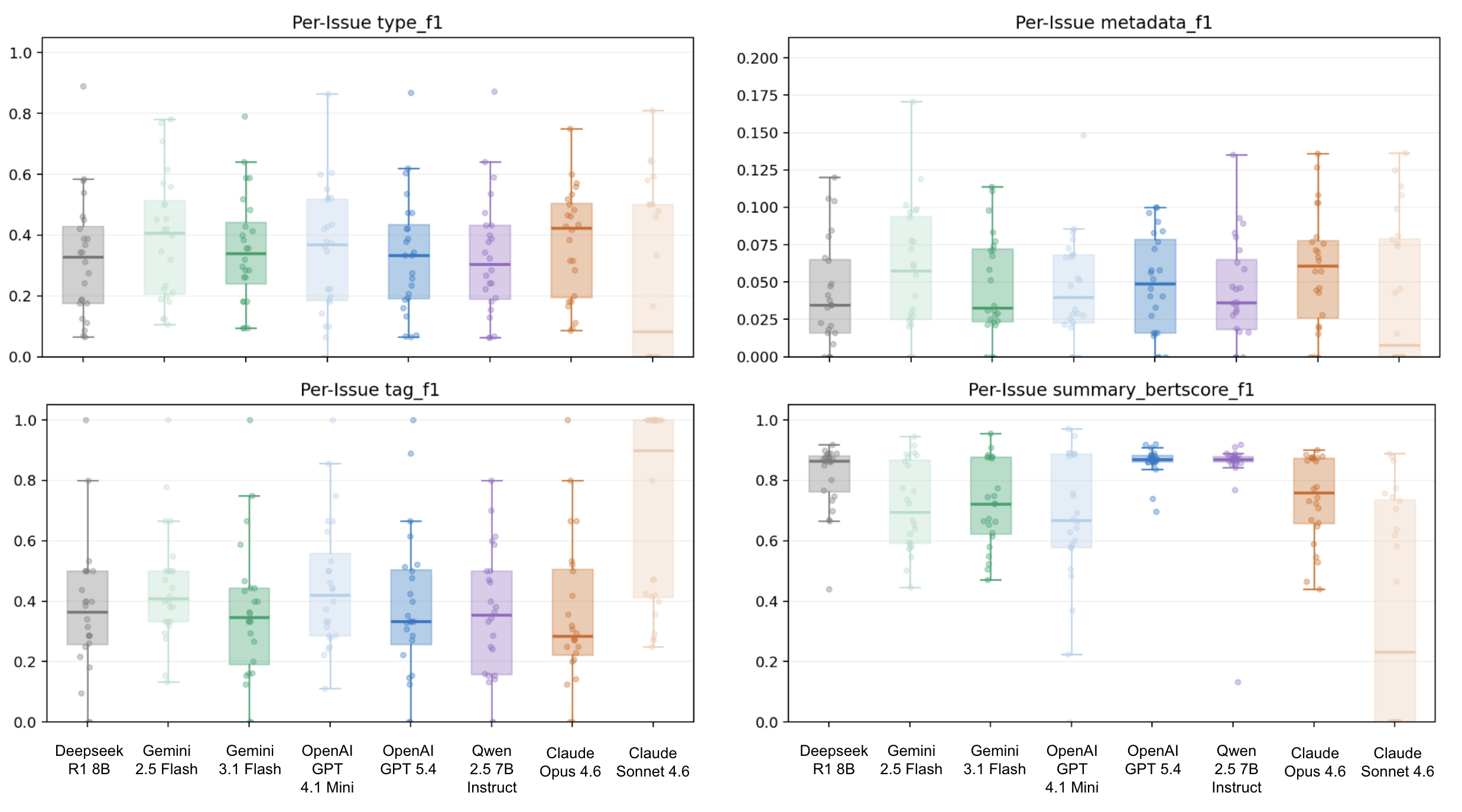}
    \caption{Box Plot for F1 scores in Type, Metadata and Tag matching for Kubernetes repository. Other repos show similar box plots. The exceptionally low metadata scores indicate that the current state of the art models are not good at discussing the right sub-topics within the type of issue. Type and Tag F1 scores are similar across models but still exhibit a scope for improvement (F1 score below 0.4). Acrtoss summary metrics, OpenAI, Qwen, and Deepseek models show a low standard deviation showing a consistent reasoning capability across different issues.}
    \label{fig:placeholder}
\end{figure}

\subsection{Global Leaderboard and Reasoning Capability}
The overall performance across all repositories and issues is summarized in the leaderboard presented in Table 3. GPT-5.4 achieved the highest overall score of 0.436, followed by DeepSeek-R1 8B at 0.422 and GPT-5 at 0.411. The most striking observation is the consistently low scores across all LLMs; even the top-performing models failed to exceed a score of 0.45. This suggests a significant gap between the capabilities of frontier models in code generation and their ability to reason through the complex design choices, constraints, and trade-offs inherent in real-world software engineering.

\begin{table}[h]
\centering
\smfoot
\caption{Global metrics and overall leaderboard score. Overall Score is calculated by averaging over all metrics (equal weight) and normalizing the score out of 100. Standard Deviation indicated in parenthesis.}
\vspace{2mm}
\label{tab:overall}
{\setlength{\tabcolsep}{3pt}%
\resizebox{\linewidth}{!}{%
\begin{tabular}{l*{7}{S[table-format=1.2]}S[table-format=1.3]}
\toprule
Model & {Meta F1} & {Meta Soft F1} & {Tag F1} & {Type F1} & {BLEURT} & {BERTScore F1} & {CodeBERT} & {Overall} \\
\midrule
\textbf{GPT-5.4} & \multicolumn{1}{c}{0.05 (0.01)} & \multicolumn{1}{c}{0.14 (0.01)} & \multicolumn{1}{c}{0.31 (0.05)} & \multicolumn{1}{c}{0.33 (0.00)} & \multicolumn{1}{c}{{\bfseries 0.39} (0.00)} & \multicolumn{1}{c}{{\bfseries 0.86} (0.00)} & \multicolumn{1}{c}{{\bfseries 0.97} (0.00)} & \multicolumn{1}{c}{{\bfseries 0.436} (0.008)} \\
GPT-5 & \multicolumn{1}{c}{0.05 (0.00)} & \multicolumn{1}{c}{0.14 (0.00)} & \multicolumn{1}{c}{0.23 (0.00)} & \multicolumn{1}{c}{0.34 (0.00)} & \multicolumn{1}{c}{0.37 (0.00)} & \multicolumn{1}{c}{0.82 (0.00)} & \multicolumn{1}{c}{0.92 (0.00)} & \multicolumn{1}{c}{0.411 (0.000)} \\
GPT-4.1 Mini & \multicolumn{1}{c}{0.05 (0.00)} & \multicolumn{1}{c}{0.14 (0.00)} & \multicolumn{1}{c}{0.33 (0.06)} & \multicolumn{1}{c}{0.34 (0.02)} & \multicolumn{1}{c}{0.29 (0.03)} & \multicolumn{1}{c}{0.63 (0.05)} & \multicolumn{1}{c}{0.70 (0.05)} & \multicolumn{1}{c}{0.353 (0.030)} \\
\midrule
Gemini 3.1 Flash & \multicolumn{1}{c}{0.05 (0.01)} & \multicolumn{1}{c}{0.15 (0.01)} & \multicolumn{1}{c}{0.31 (0.03)} & \multicolumn{1}{c}{0.36 (0.01)} & \multicolumn{1}{c}{0.33 (0.00)} & \multicolumn{1}{c}{0.72 (0.00)} & \multicolumn{1}{c}{0.81 (0.00)} & \multicolumn{1}{c}{0.391 (0.002)} \\
Gemini 2.5 Flash & \multicolumn{1}{c}{{\bfseries 0.06} (0.00)} & \multicolumn{1}{c}{{\bfseries 0.17} (0.00)} & \multicolumn{1}{c}{{\bfseries 0.36} (0.08)} & \multicolumn{1}{c}{{\bfseries 0.37} (0.03)} & \multicolumn{1}{c}{0.32 (0.03)} & \multicolumn{1}{c}{0.68 (0.05)} & \multicolumn{1}{c}{0.76 (0.05)} & \multicolumn{1}{c}{0.386 (0.034)} \\
\midrule
Claude Opus 4.6 & \multicolumn{1}{c}{{\bfseries 0.06} (0.00)} & \multicolumn{1}{c}{{\bfseries 0.17} (0.00)} & \multicolumn{1}{c}{0.35 (0.02)} & \multicolumn{1}{c}{0.35 (0.04)} & \multicolumn{1}{c}{0.31 (0.05)} & \multicolumn{1}{c}{0.67 (0.09)} & \multicolumn{1}{c}{0.76 (0.10)} & \multicolumn{1}{c}{0.383 (0.038)} \\
Claude Sonnet 4.6 & \multicolumn{1}{c}{0.05 (0.02)} & \multicolumn{1}{c}{0.15 (0.04)} & \multicolumn{1}{c}{0.35 (0.05)} & \multicolumn{1}{c}{0.30 (0.06)} & \multicolumn{1}{c}{0.24 (0.10)} & \multicolumn{1}{c}{0.52 (0.22)} & \multicolumn{1}{c}{0.58 (0.25)} & \multicolumn{1}{c}{0.312 (0.091)} \\
\midrule
\textbf{DeepSeek-R1 8B} & \multicolumn{1}{c}{0.05 (0.01)} & \multicolumn{1}{c}{0.13 (0.00)} & \multicolumn{1}{c}{0.32 (0.06)} & \multicolumn{1}{c}{0.33 (0.01)} & \multicolumn{1}{c}{{\bfseries 0.39} (0.01)} & \multicolumn{1}{c}{0.83 (0.02)} & \multicolumn{1}{c}{0.92 (0.02)} & \multicolumn{1}{c}{{\bfseries 0.422} (0.000)} \\
Qwen2.5 7B & \multicolumn{1}{c}{0.04 (0.01)} & \multicolumn{1}{c}{0.12 (0.01)} & \multicolumn{1}{c}{0.31 (0.07)} & \multicolumn{1}{c}{0.30 (0.04)} & \multicolumn{1}{c}{0.35 (0.08)} & \multicolumn{1}{c}{0.73 (0.16)} & \multicolumn{1}{c}{0.81 (0.18)} & \multicolumn{1}{c}{0.381 (0.078)} \\
\bottomrule
\end{tabular}%
}}
\end{table}

\subsection{Baseline and Ablation Analysis}
To quantify the impact of the various components in the reasoning pipeline, an extensive ablation study was performed across four primary configurations: the proposed hybrid RAG pipeline (\textit{Orig.}), a \textit{Few-Shot} setting, a \textit{No Context} (NoCtx) setting, and a sparse retrieval baseline (\textit{BM25 Only}). The results, detailed in Table 4, Table 5 and Figure 6, underscore that while specific prompting or retrieval strategies can shift metrics, current LLMs still exhibit a fundamental struggle with deep repository-level reasoning.

\begin{itemize}
    \item \textbf{Contextual and Retrieval Ablations:} The catastrophic performance of the \textit{No Context} baseline confirms that engineering reasoning is highly dependent on grounded technical facts. In this setting, \texttt{Type F1} scores collapsed to a baseline average of $0.11$ across all models, and semantic similarity measures like BERTScore fell into the $0.15$--$0.22$ range. While the \textit{BM25 Only} baseline occasionally yielded higher \texttt{Tag F1} and \texttt{Type F1} scores—reaching $0.50$ for Gemini 2.5 Flash—it frequently struggled with semantic alignment compared to the hybrid dense-sparse architecture. This suggests that while keyword matching can help identify the "topic" of a discussion, it is insufficient for replicating the actual logic and deliberative content found in human engineering threads.

    \item \textbf{Impact of Few-Shot Prompting:} The inclusion of repository-matched few-shot examples generally provided a performance boost in structured identification tasks. For instance, Gemini 2.5 Flash saw its \texttt{Tag F1} increase from $0.36$ in the \textit{Orig.} setting to $0.42$ with few-shot guidance. However, this gain did not consistently translate to semantic summary quality, as metrics like BLEURT remained relatively stagnant or showed only marginal improvements. This indicates that while models can be steered toward the correct taxonomic labels through in-context learning, the underlying "thought" generation—specifically regarding \texttt{trade\_off\_argument}s and \texttt{design\_decision}s—remains limited by the model's intrinsic reasoning capabilities.

    \item \textbf{Heuristic vs. Neural Extraction:} The extraction pipeline was further evaluated against a \textit{Regex-based} heuristic baseline. As illustrated in the global average tag frequency analysis (Figure 7, Appendix-B), the Regex approach is significantly more "noisy," frequently over-tagging generic technical terms without regard for context. For example, mentions of "test" were universally tagged as \texttt{unit\_test} artifacts, leading to inflated recall but extremely poor precision. In contrast, the neural extraction system employed in the benchmark demonstrates a more nuanced understanding of how specific technical artifacts function within a deliberation, though it still faces a significant challenge in capturing the complex logic required for accurate \texttt{metadata} extraction.
\end{itemize}

\begin{figure}[H]
    \centering
    \includegraphics[width=0.9\linewidth]{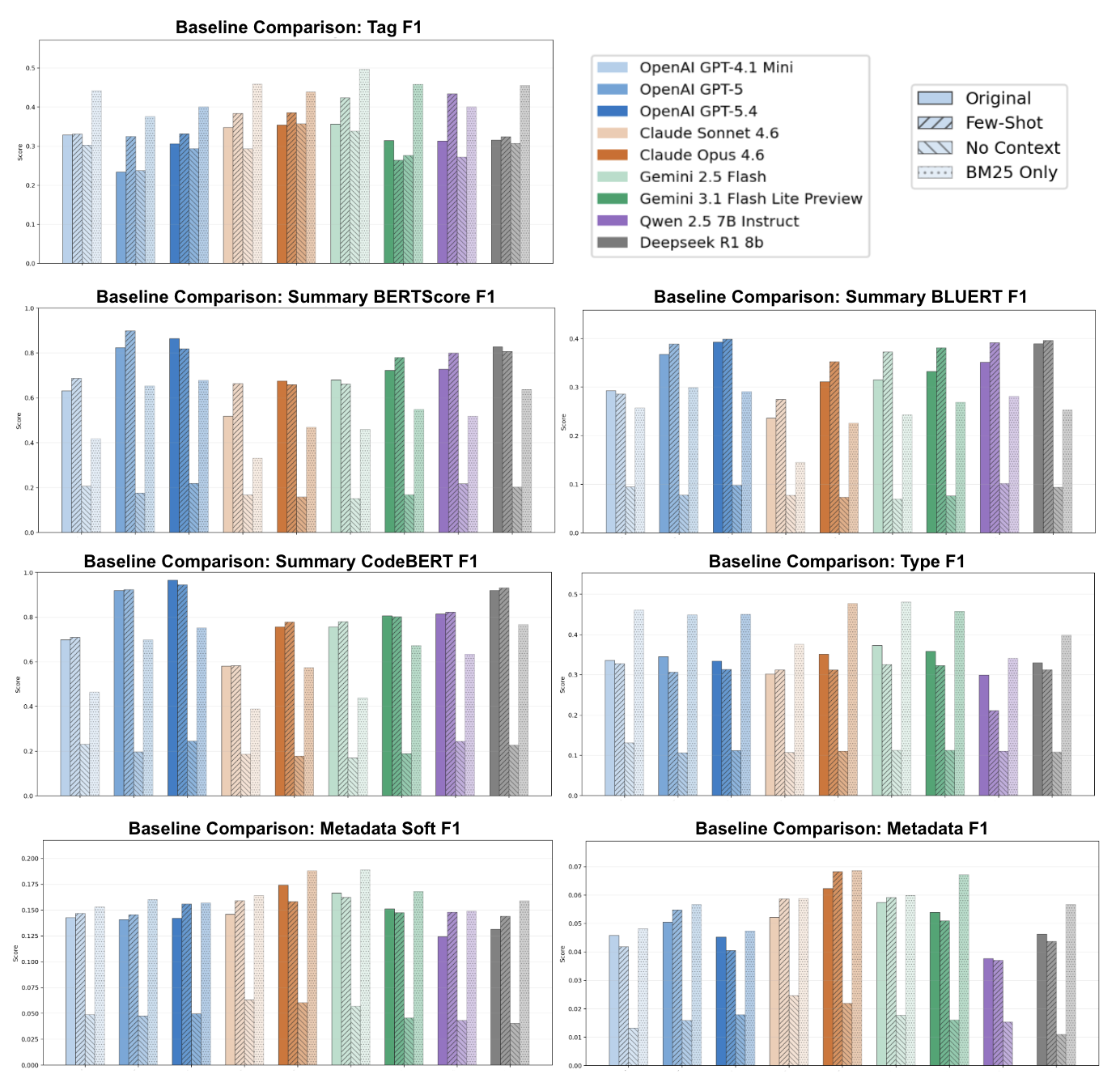}
    \caption{Baseline Comparisons and Ablation Study: The figure shows all metrics on the Kubernetes and Bitcoin issues. Adding few-shots to the prompt in Phase 5 leads to slightly better tagging. However, the additional cost of providing few shots within the prompt outweigh the benefit. Excluding the code context from the prompt significantly reduces performance across metrics. Proving code context using only BM25 as the RAG component leads to better tagging and type matching, which comes at the cost of a poor summary. Because of these the `Original' system uses BM25 + Dense Retrieval System (combined with Reciprocal Rank Fusion) without few-shots to keep the system effective and cost efficient.}
    \label{fig:placeholder}
\end{figure}

\begin{table}[H]
\centering
\footnotesize
\setlength{\tabcolsep}{4pt}
\caption{Baseline comparison (Metadata, Tag, Type). Few (Few Shot prompt), NoCtx (No code context in the prompt), BM25 (only BM25 RAG, no dense retrieval), Orig. (BM25+ Dense Retrieval code context without few shot prompts).}
\vspace{1mm}
\label{tab:baselines_part1}
\begin{tabular}{l*{12}{S[table-format=1.2]}}
\toprule
\multirow{2}{*}{Model}
& \multicolumn{4}{c}{Metadata Soft F1}
& \multicolumn{4}{c}{Tag F1}
& \multicolumn{4}{c}{Type F1} \\
\cmidrule(lr){2-5}
\cmidrule(lr){6-9}
\cmidrule(lr){10-13}
& \multicolumn{1}{c}{Orig.} & \multicolumn{1}{c}{Few} & \multicolumn{1}{c}{NoCtx} & \multicolumn{1}{c}{BM25}
& \multicolumn{1}{c}{Orig.} & \multicolumn{1}{c}{Few} & \multicolumn{1}{c}{NoCtx} & \multicolumn{1}{c}{BM25}
& \multicolumn{1}{c}{Orig.} & \multicolumn{1}{c}{Few} & \multicolumn{1}{c}{NoCtx} & \multicolumn{1}{c}{BM25} \\
\midrule
GPT-5.4 & 0.14 & {\bfseries 0.16} & 0.05 & {\bfseries 0.16} & 0.31 & 0.33 & 0.29 & {\bfseries 0.40} & 0.33 & 0.31 & 0.11 & {\bfseries 0.45} \\
GPT-5 & 0.14 & 0.15 & 0.05 & {\bfseries 0.16} & 0.23 & 0.33 & 0.24 & {\bfseries 0.38} & 0.34 & 0.31 & 0.11 & {\bfseries 0.45} \\
GPT-4.1 Mini & 0.14 & {\bfseries 0.15} & 0.05 & {\bfseries 0.15} & 0.33 & 0.33 & 0.30 & {\bfseries 0.44} & 0.34 & 0.33 & 0.13 & {\bfseries 0.46} \\
\midrule
Gemini 3.1 Flash & 0.15 & 0.15 & 0.05 & {\bfseries 0.17} & 0.31 & 0.26 & 0.28 & {\bfseries 0.46} & 0.36 & 0.32 & 0.11 & {\bfseries 0.46} \\
Gemini 2.5 Flash & 0.17 & 0.16 & 0.06 & {\bfseries 0.19} & 0.36 & 0.42 & 0.34 & {\bfseries 0.50} & 0.37 & 0.33 & 0.11 & {\bfseries 0.48} \\
\midrule
Claude Opus 4.6 & 0.17 & 0.16 & 0.06 & {\bfseries 0.19} & 0.35 & 0.39 & 0.36 & {\bfseries 0.44} & 0.35 & 0.31 & 0.11 & {\bfseries 0.48} \\
Claude Sonnet 4.6 & 0.15 & {\bfseries 0.16} & 0.06 & {\bfseries 0.16} & 0.35 & 0.38 & 0.29 & {\bfseries 0.46} & 0.30 & 0.31 & 0.11 & {\bfseries 0.38} \\
\midrule
DeepSeek-R1 8B & 0.13 & 0.14 & 0.04 & {\bfseries 0.16} & 0.32 & 0.32 & 0.31 & {\bfseries 0.46} & 0.33 & 0.31 & 0.11 & {\bfseries 0.40} \\
Qwen2.5 7B & 0.12 & {\bfseries 0.15} & 0.04 & {\bfseries 0.15} & 0.31 & {\bfseries 0.43} & 0.27 & 0.40 & 0.30 & 0.21 & 0.11 & {\bfseries 0.34} \\
\bottomrule
\end{tabular}
\end{table}

\begin{table}[H]
\centering
\footnotesize
\setlength{\tabcolsep}{4pt}
\caption{Baseline comparison - Summary (BLEURT, BERTScore, CodeBERT). Few (Few Shot prompt), NoCtx (No code context in the prompt), BM25 (only BM25 RAG, no dense retrieval), Orig. (BM25+ Dense Retrieval code context without few shot prompts).}
\vspace{1mm}
\label{tab:baselines_part2}
\begin{tabular}{l*{12}{S[table-format=1.2]}}
\toprule
\multirow{2}{*}{Model}
& \multicolumn{4}{c}{BLEURT}
& \multicolumn{4}{c}{BERTScore F1}
& \multicolumn{4}{c}{CodeBERT} \\
\cmidrule(lr){2-5}
\cmidrule(lr){6-9}
\cmidrule(lr){10-13}
& \multicolumn{1}{c}{Orig.} & \multicolumn{1}{c}{Few} & \multicolumn{1}{c}{NoCtx} & \multicolumn{1}{c}{BM25}
& \multicolumn{1}{c}{Orig.} & \multicolumn{1}{c}{Few} & \multicolumn{1}{c}{NoCtx} & \multicolumn{1}{c}{BM25}
& \multicolumn{1}{c}{Orig.} & \multicolumn{1}{c}{Few} & \multicolumn{1}{c}{NoCtx} & \multicolumn{1}{c}{BM25} \\
\midrule
GPT-5.4 & 0.39 & {\bfseries 0.40} & 0.10 & 0.29 & {\bfseries 0.86} & 0.82 & 0.22 & 0.68 & {\bfseries 0.97} & 0.94 & 0.24 & 0.75 \\
GPT-5 & 0.37 & {\bfseries 0.39} & 0.08 & 0.30 & 0.82 & {\bfseries 0.90} & 0.18 & 0.65 & {\bfseries 0.92} & {\bfseries 0.92} & 0.20 & 0.70 \\
GPT-4.1 Mini & {\bfseries 0.29} & {\bfseries 0.29} & 0.09 & 0.26 & 0.63 & {\bfseries 0.69} & 0.21 & 0.42 & 0.70 & {\bfseries 0.71} & 0.23 & 0.46 \\
\midrule
Gemini 3.1 Flash & 0.33 & {\bfseries 0.38} & 0.08 & 0.27 & 0.72 & {\bfseries 0.78} & 0.17 & 0.55 & {\bfseries 0.81} & 0.80 & 0.19 & 0.67 \\
Gemini 2.5 Flash & 0.32 & {\bfseries 0.37} & 0.07 & 0.24 & {\bfseries 0.68} & 0.66 & 0.15 & 0.46 & 0.76 & {\bfseries 0.78} & 0.17 & 0.44 \\
\midrule
Claude Opus 4.6 & 0.31 & {\bfseries 0.35} & 0.07 & 0.23 & {\bfseries 0.67} & 0.66 & 0.16 & 0.47 & 0.76 & {\bfseries 0.78} & 0.18 & 0.57 \\
Claude Sonnet 4.6 & 0.24 & {\bfseries 0.27} & 0.08 & 0.15 & 0.52 & {\bfseries 0.66} & 0.17 & 0.33 & {\bfseries 0.58} & {\bfseries 0.58} & 0.19 & 0.39 \\
\midrule
DeepSeek-R1 8B & 0.39 & {\bfseries 0.40} & 0.09 & 0.25 & {\bfseries 0.83} & 0.81 & 0.20 & 0.64 & 0.92 & {\bfseries 0.93} & 0.23 & 0.77 \\
Qwen2.5 7B & 0.35 & {\bfseries 0.39} & 0.10 & 0.28 & 0.73 & {\bfseries 0.80} & 0.22 & 0.52 & 0.81 & {\bfseries 0.82} & 0.24 & 0.64 \\
\bottomrule
\end{tabular}
\end{table}

\section{Conclusion}
\vspace{-5mm}

To summarize, the following has been achieved by this study:

\begin{enumerate}
\item \textbf{Refined Artifact Extraction System:} This goal has been fully operationalized through a concrete schema and a validated runtime pipeline. The extraction system defines a comprehensive artifact space, requiring models to adhere to fixed metadata fields and a constrained tag vocabulary. The implementation utilizes Pydantic for output validation and records extraction failures on a per-comment basis to ensure data integrity. To date, the system has successfully curated a large dataset as described above.

\item \textbf{Prompting Methods Implemented and Tested:} A full prompting loop was implemented that integrates hybrid retrieval (BM25 + CodeBERT with reciprocal-rank fusion) and structured deliberation traces. The ``structured deliberation" is operationalized through a thought prompt that mandates topic-labeled engineering comments. Furthermore, the system supports few-shot prompting by dynamically loading prior human extraction examples into the inference context. The custom CLI facilitates the original RAG, few-shot, no-context, and heuristic baselines, allowing for the comprehensive ablation studies presented in Section 5.

\item \textbf{Evaluation Framework:} The framework computes alignment across four dimensions: type matching, metadata matching, tag matching, and semantic summary similarity. The orchestration code supports batch execution of metrics and generates a combined leaderboard using rank-fusion. The metrics such as "decision agreement" are currently studied thoroughly and the visualization suite effectively translates these raw outputs into comparative graphs and tables.

\item \textbf{Comprehensive Benchmark Results and Analysis:} The project has delivered a complete set of benchmark outputs for nine frontier model families. These results include a global leaderboard where GPT-5.4 currently ranks as the top performer. To facilitate qualitative analysis, I developed a side-by-side viewer that assembles raw GitHub issues, human "gold" extractions, and model-derived artifacts within a unified interface. This enables researchers to perform deep-dive audits of model reasoning quality across multiple repositories.
\end{enumerate}

\bibliographystyle{plainnat}
\bibliography{references}

\newpage

\appendix
\begin{center}
    \huge\bfseries APPENDIX
\end{center}

\section{Prompt Templates}

Following is the prompt given to \texttt{gemma3:12b}, \texttt{gemma4:e4b}, \textbf{llama3.1:8b} to extract the artifacts from the human comments in the GitHub issues and from the thought responses of the SoTA LLMs:
\begin{mytbox}
\begin{verbatim}
<System Prompt>

Extract discussion artifacts from one GitHub comment.
Return strict JSON only: {"artifacts": [...]}.

Rules:
1) type: required, one of [<allowed_artifact_types>].
2) summary: required, specific key takeaway (not generic). Include
   - subject (component/file/feature)
   - event (bug/change/question/suggestion)
   - key technical detail (error/behavior/number/condition/cause if known)
3) tags: required, at least 1 tag, use as many as needed, unique, lowercase snake_case, 
         only from [<allowed_tags>].
   Be generous with relevant tags: include all tags clearly supported by the comment.
   For each candidate tag, explicitly check if evidence in the comment supports it; add
   it when supported, skip when not supported.
   Prefer broader correct coverage over minimal tagging, but do not add speculative tags.
4) metadata: required; include exactly the fixed keys for the chosen type, never use 
             null, no extra keys.
5) Use only the provided comment text; no outside knowledge.
6) If no valid artifact exists, return {"artifacts": []}.

<User Prompt>

Repo: <repository>
Issue: <issue_number>
Comment JSON:
{"author":"<comment_author>","link":"<comment_link>","text":"<comment_text>"}

Metadata keys by type (use exact keys; never null; use 'unknown' when needed):
- <artifact_type_1>: <metadata_key_1>, <metadata_key_2>, ...
- <artifact_type_2>: <metadata_key_1>, <metadata_key_2>, ...
- ...
\end{verbatim}
\end{mytbox}

\newpage
Following is the prompt given to the LLMs for thinking like human engineers to produce their artifacts:

\begin{mytbox}
\begin{verbatim}
You are given a GitHub issue. Think like a human engineer reading the issue 
for the first time.

Instructions:
- Produce at least 20 short comments
- Each comment must be on its own line
- Each line must follow the format: [Topic] comment
- Comments may be conflicting, uncertain, speculative, or partially redundant
- Use only the following topics: <allowed_thought_topics>

Issue:
- Number: <issue_number>
- Title: <issue_title>
- Body: <issue_body>

FEW SHOT EXAMPLES:
<optional_few_shot_examples>

CODE SNIPPETS:
<file_1.py>
<retrieved_code_context>
....
<file_n.py>
<retrieved_code_context>

\end{verbatim}
\end{mytbox}

\section{Dataset Details}

\begin{figure}[H]
    \centering
    \includegraphics[width=0.8\linewidth]{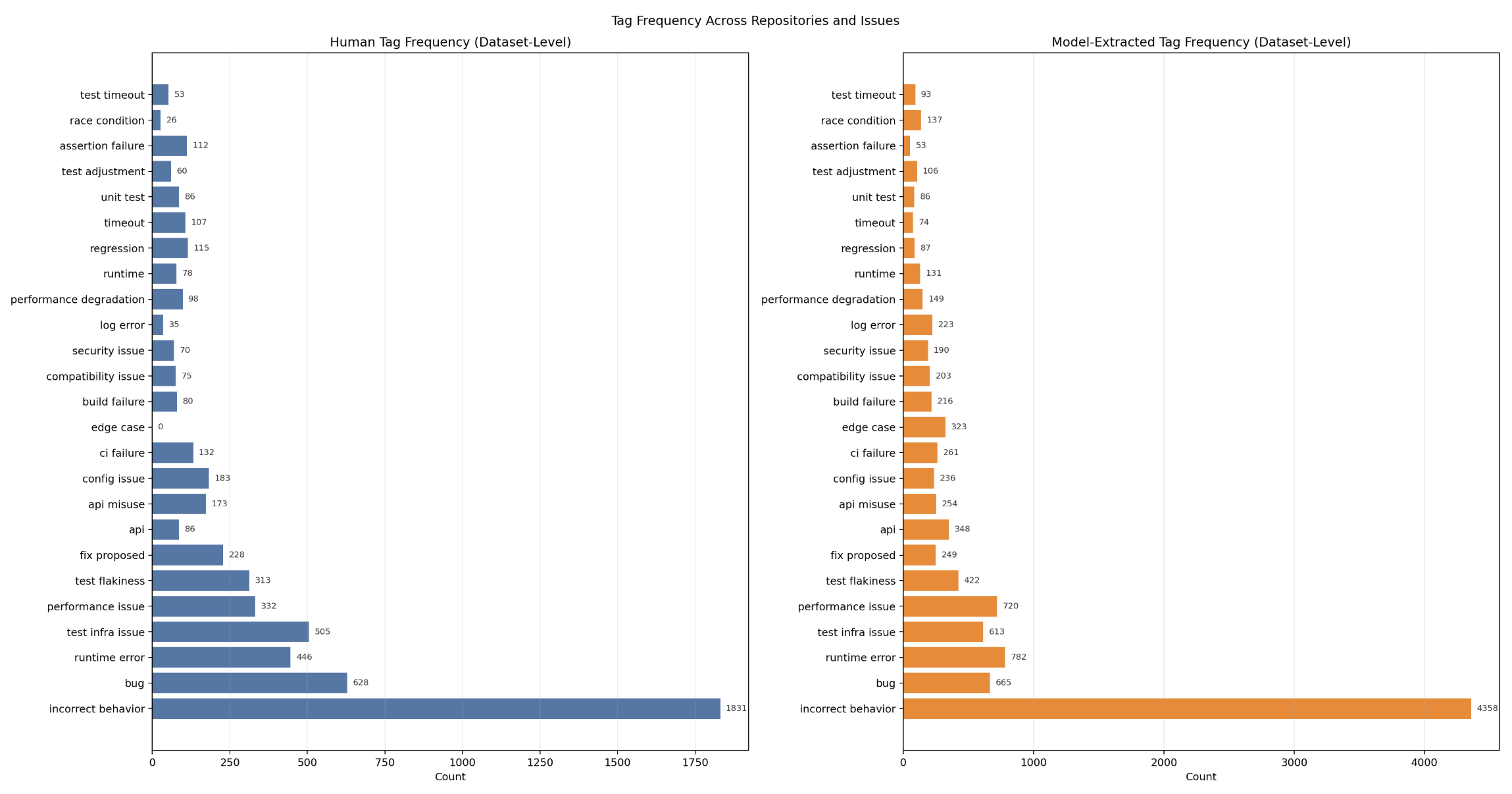}
    \caption{Tags extracted from GitHub comments (left) and LLM responses (right). While the distribution is similar, there are subtle differences as well. For example, the human comments tend to have more discussion around bugs whereas LLMs treat these as edge cases because they are conditioned to view errors as ``missing test cases" rather than actual design issues.}
    \label{fig:placeholder}
\end{figure}

Allowed tag categories:
\begin{mytbox}
\tiny
\begin{verbatim}
bug, regression, flaky_test, ci_failure, build_failure, runtime_error, compile_error,assertion_failure, performance_issue, memory_issue, security_issue,
data_corruption, incorrect_behavior, edge_case, compatibility_issue, race_condition, non_determinism, timestamp_issue, floating_point_error, off_by_one, 
null_pointer, uninitialized_state, state_sync_issue, logic_bug, api_misuse, config_issue, dependency_issue, version_mismatch, platform_specific, 
overflow_underflow, unit_test, integration_test, e2e_test, functional_test, test_framework_issue, test_flakiness, test_timeout, test_infra_issue, 
networking, filesystem, database, api, cli, ui, backend, frontend, distributed_system, p2p, consensus, mempool, rpc, serialization, crypto, validation, 
scheduler, training, inference, data_pipeline, tokenization, linux, macos, windows, arm, x86, docker, ci_environment, gpu, low_memory, build_time, 
startup, runtime, shutdown, deployment, crash, hang, timeout, incorrect_output, missing_output, extra_output, log_error,performance_degradation, 
workaround, fix_proposed, fix_merged, refactor, test_adjustment, relax_assertion.    
\end{verbatim}
\end{mytbox}

\begin{table}[H]
\centering
\caption{Entity classes used for Extraction.}
\vspace{2mm}
\begin{tabular}{lp{10.2cm}}
\hline
\hline
\textbf{Annotation Type} & \textbf{Description} \\
\hline
\hline
Problem Statement & Description of a bug, feature request, or user need. \\
\hline
Proposed Solution & Suggested fix, implementation, or approach. \\
\hline
Alternative Solution & Competing or different approach to solve the same problem. \\
\hline
Design Decision & Final or tentative choice among alternatives. \\
\hline
Trade-off Argument & Comparison between options highlighting pros and cons. \\
\hline
Rationale & Explanation or justification for a decision. \\
\hline
Constraint & Limitation or requirement such as performance or compatibility. \\
\hline
Assumption & Stated or implicit premise about the system or inputs. \\
\hline
Implementation Detail & Low-level description of how something should be built. \\
\hline
Code Snippet & Concrete code example or patch. \\
\hline
Algorithm / Approach & High-level method or technique. \\
\hline
API Design & Discussion of interfaces, endpoints, or function signatures. \\
\hline
Data Structure Choice & Selection of data structures and their implications. \\
\hline
Configuration Choice & Decisions about parameters or settings. \\
\hline
Benchmark Result & Measured performance or experimental result. \\
\hline
Performance Claim & Assertion about speed, memory, or efficiency. \\
\hline
Test Case & Example used to validate correctness. \\
\hline
Bug Reproduction Steps & Steps to reproduce an issue. \\
\hline
Edge Case & Rare or boundary scenario that needs handling. \\
\hline
Empirical Evidence & Data or experiments supporting a claim. \\
\hline
Question & Inquiry asking for clarification or information. \\
\hline
Answer / Clarification & Response resolving a question. \\
\hline
Agreement & Expression of alignment or approval. \\
\hline
Disagreement & Expression of conflict or opposing view. \\
\hline
Suggestion & Recommendation for improvement. \\
\hline
Critique & Evaluation or review of a proposal. \\
\hline
Task Assignment & Delegation of work to a contributor. \\
\hline
Status Update & Progress or current state of work. \\
\hline
Priority Discussion & Conversation about importance or urgency. \\
\hline
Blocking Issue & Issue preventing further progress. \\
\hline
Dependency & Relationship to another issue, module, or system. \\
\hline
\hline
\end{tabular}
\label{tab:discussion-entity-schema}
\end{table}

\begin{table}[H]
\label{tab:artifact-metadata-fields}
\caption{Metadata fields required for each artifact type in the discussion-artifact extraction schema.}
\vspace{1mm}
\centering
\begin{tabular}{p{0.28\textwidth} p{0.66\textwidth}}
\hline
\hline
\textbf{Artifact Type} & \textbf{Metadata Fields} \\
\hline
\hline
problem\_statement & problem\_kind, affected\_area, user\_or\_system\_impact \\
\hline
proposed\_solution & solution\_kind, target\_area, expected\_benefit \\
\hline
alternative\_solution & alternative\_to, differentiator, comparison\_focus \\
\hline
design\_decision & decision\_status, chosen\_option, rejected\_options \\
\hline
trade\_off\_argument & options\_compared, comparison\_axes, preferred\_option \\
\hline
rationale & supports\_artifact\_type, reasoning\_kind, impact\_scope \\
\hline
constraint & constraint\_type, scope, severity \\
\hline
assumption & assumption\_scope, confidence, failure\_mode \\
\hline
implementation\_detail & component, lifecycle\_stage, implementation\_action \\
\hline
code\_snippet & language, snippet\_kind, intended\_use \\
\hline
algorithm\_approach & approach\_name, target\_problem, scale\_or\_complexity\_notes \\
\hline
api\_design & interface\_kind, api\_surface, compatibility\_impact \\
\hline
data\_structure\_choice & data\_structure, workload, motivation \\
\hline
configuration\_choice & parameter, selected\_value, config\_scope \\
\hline
benchmark\_result & metric, baseline, candidate, environment \\
\hline
performance\_claim & metric\_focus, claim\_direction, workload \\
\hline
test\_case & test\_level, coverage\_target, expected\_outcome \\
\hline
bug\_reproduction\_steps & environment, prerequisites, determinism \\
\hline
edge\_case & trigger, impact, rarity \\
\hline
empirical\_evidence & evidence\_type, sample\_source, reliability \\
\hline
question & question\_kind, target\_area, urgency \\
\hline
answer\_clarification & target\_question\_topic, certainty, resolution\_status \\
\hline
agreement & target, strength, justification \\
\hline
disagreement & target, contention\_point, severity \\
\hline
suggestion & target\_area, expected\_improvement, urgency \\
\hline
critique & target, critique\_dimension, severity \\
\hline
task\_assignment & assignee, deliverable, due\_context \\
\hline
status\_update & status\_kind, progress\_state, blocking \\
\hline
priority\_discussion & priority\_level, compared\_against, scope \\
\hline
blocking\_issue & blocker\_type, blocking\_target, unblock\_condition \\
\hline
dependency & dependency\_type, dependency\_target, relationship \\
\hline
\hline
\end{tabular}
\end{table}

\section{Similarity Metrics Used}
For normalized phrases $a$ and $b$, let $T(a)$ be the set of whitespace-delimited tokens, and let $C_n(a)$ be the set of character $n$-grams of the padded string ``\verb| |\,$a$\,\verb| |''.

\textbf{Token F1:}
Let $o = |T(a)\cap T(b)|$. Then:
\[
\operatorname{tokPrec}(a,b)=\operatorname{ratio}(o,|T(b)|),\qquad
\operatorname{tokRec}(a,b)=\operatorname{ratio}(o,|T(a)|),
\]
\[
\operatorname{tokenF1}(a,b)=\operatorname{harm}(\operatorname{tokPrec}(a,b),\operatorname{tokRec}(a,b)).
\]

Measures overlap between unique token sets.

\textbf{Sequence Ratio:}
\(
\operatorname{sequenceRatio}(a,b)=\texttt{SequenceMatcher}(a,b).\texttt{ratio()}.
\)

Measures surface-form similarity of the normalized strings.

\textbf{Token Jaccard:}
\(
\operatorname{tokenJaccard}(a,b)=\operatorname{ratio}(|T(a)\cap T(b)|,|T(a)\cup T(b)|).
\)

Measures set overlap between token vocabularies.

\textbf{Character 3-gram Jaccard:}
\(
\operatorname{char3Jaccard}(a,b)=\operatorname{ratio}(|C_3(a)\cap C_3(b)|,|C_3(a)\cup C_3(b)|).
\)

Measures fuzzy overlap at the subword character level.

\textbf{Token Containment:}
\(
\operatorname{cont}(a,b)=\operatorname{ratio}(|T(a)\cap T(b)|,\min(|T(a)|,|T(b)|)).
\)

Measures whether the shorter token set is mostly contained in the longer one.

\textbf{Max-All Ensemble.}
\[
\operatorname{maxAll}(a,b)=\max\{
\operatorname{tokenF1}(a,b),
\operatorname{sequenceRatio}(a,b),
\operatorname{tokenJaccard}(a,b),
\operatorname{char3Jaccard}(a,b),
\operatorname{cont}(a,b)
\}.
\]
Measures best-case phrase agreement across several lightweight similarity functions.

\subsection{Greedy One-to-One Matching:}
For one field $f$ of one type $t$ in one issue $i$, compute all pairwise similarities:
\[
\sigma_{jk}=\operatorname{sim}(v_j,u_k),
\quad v_j \in V_{i,t,f}^{H},\ u_k \in V_{i,t,f}^{L}.
\]
Sort all candidate pairs by descending $\sigma_{jk}$ and greedily keep a pair only if neither value has been used before. This produces a one-to-one set of greedy pairs $P$.

\section{BLEURT Calculation}
Let $\beta_{\mathrm{raw}}(x,y)$ be the scalar output of the configured BLEURT sequence-classification model. The reported score depends on post-processing mode.

\textbf{Raw mode:}
\(
\beta(x,y)=\beta_{\mathrm{raw}}(x,y).
\)

\textbf{Clip mode:}
For clipping floor $c_{\min}$:
\(
\beta(x,y)=
\begin{cases}
\max(c_{\min},\beta_{\mathrm{raw}}(x,y)) & c_{\min}\neq \varnothing\\
\beta_{\mathrm{raw}}(x,y) & \text{otherwise}.
\end{cases}
\)

\textbf{Sigmoid mode:}
For temperature $T>0$ and bias $b$:
\[
z=\frac{\beta_{\mathrm{raw}}(x,y)-b}{T},
\qquad
z \leftarrow \min(60,\max(-60,z)),\qquad \beta(x,y)=\frac{1}{1+e^{-z}}
\]
The implementation returns $1$ for two empty strings and $0$ if exactly one side is empty. Off these, the sigmoid mode was deemed to be the best and used for all the experiments.

\begin{figure}[H]
    \centering
    \includegraphics[width=0.6\linewidth]{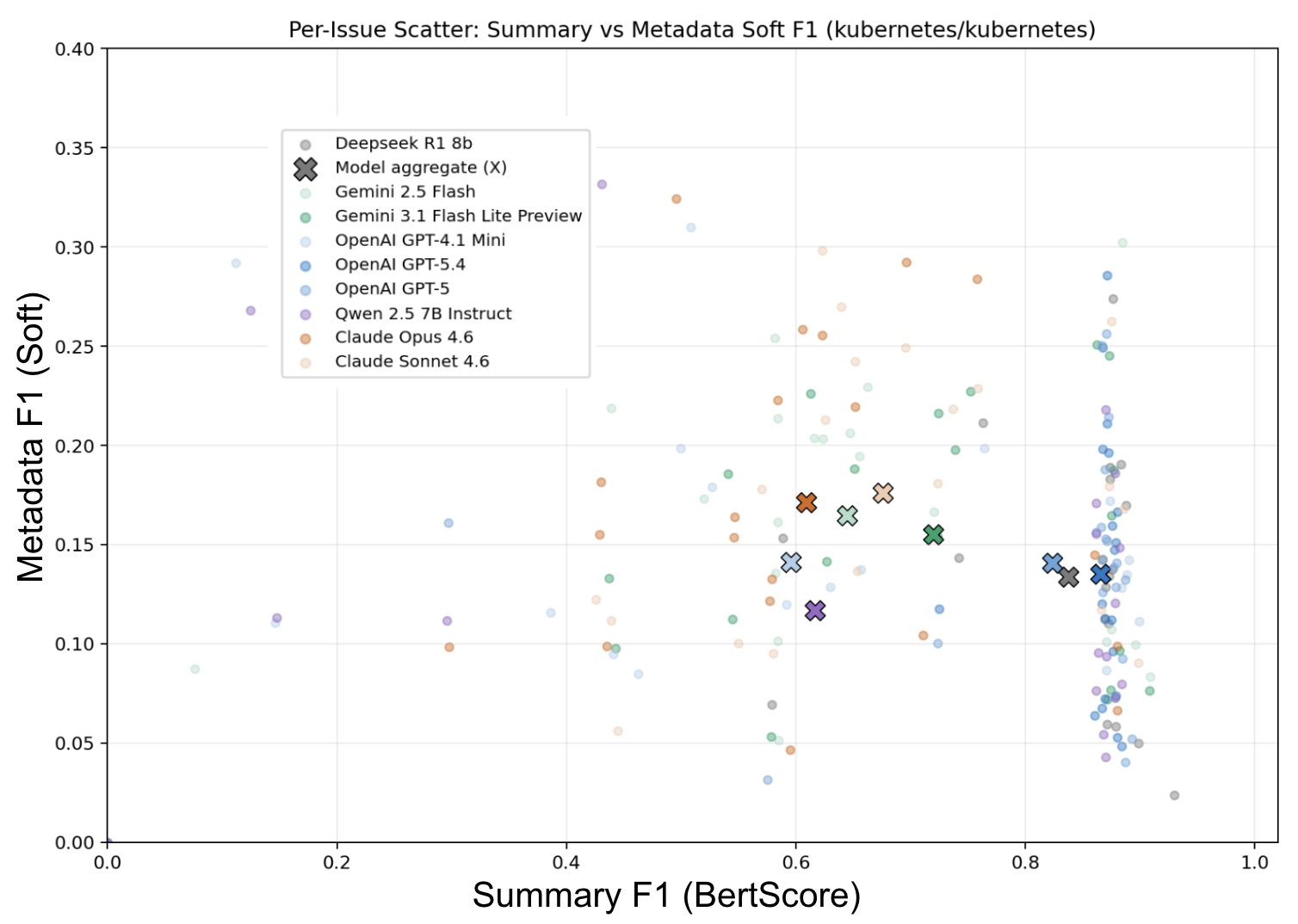}
    \caption{Correlation comparison between metadata summary metrics. The OpenAI models (blue) concentrate towards the right - showing superior summary but varying metadata accuracy across runs. No correlation appears among other model families.}
    \label{fig:placeholder}
\end{figure}

\section{Visualizer}

\begin{figure}[H]
    \centering
    \includegraphics[width=0.6\linewidth]{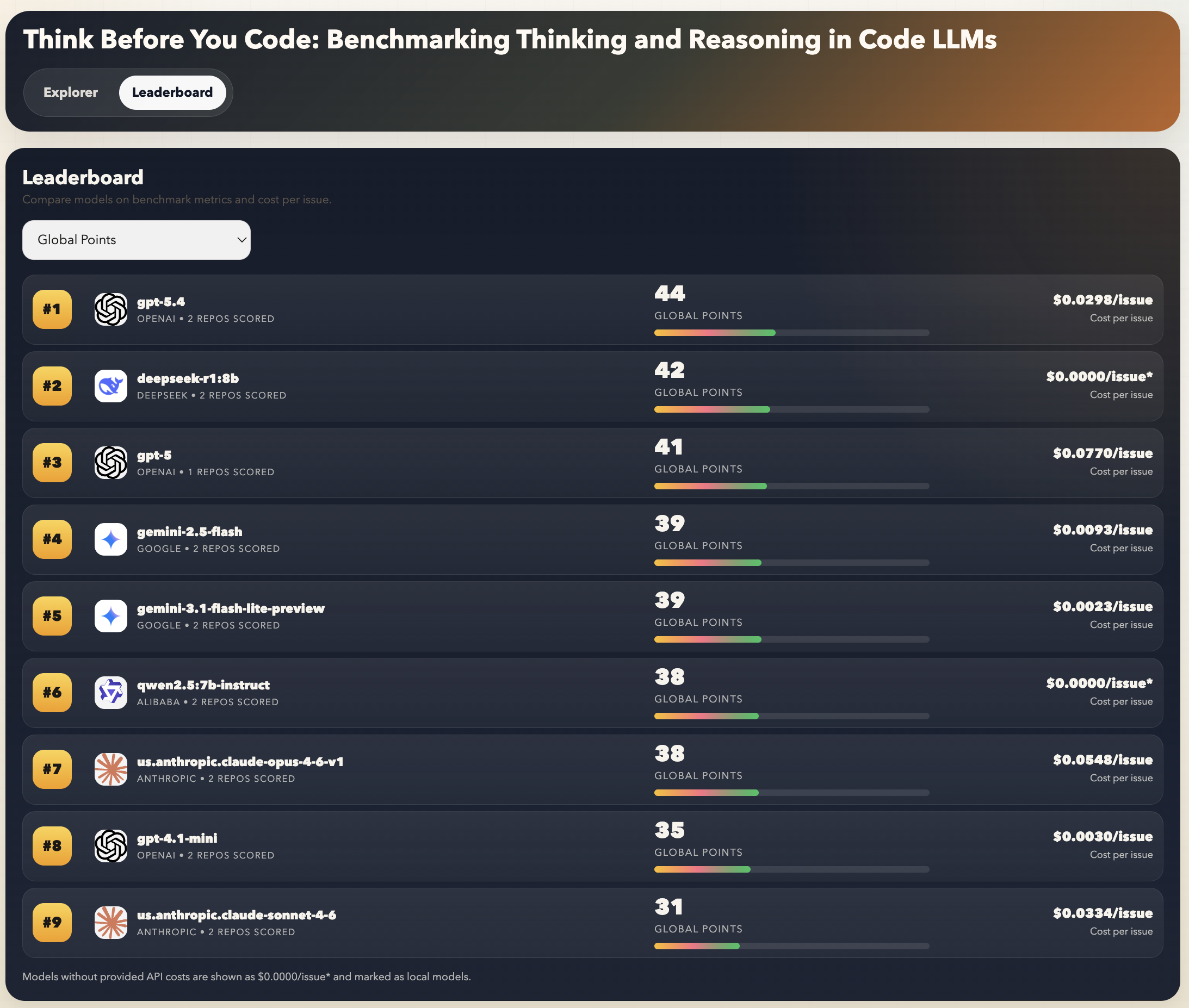}
    \caption{Leaderboard for TBYC - An interactive leader board to see the ranking and scores of models across various metrics and a global scoring schema. The global score is calculated as an average overall metrics.}
    \label{fig:placeholder}
\end{figure}

\begin{figure}[H]
    \centering
    \includegraphics[width=0.7\linewidth]{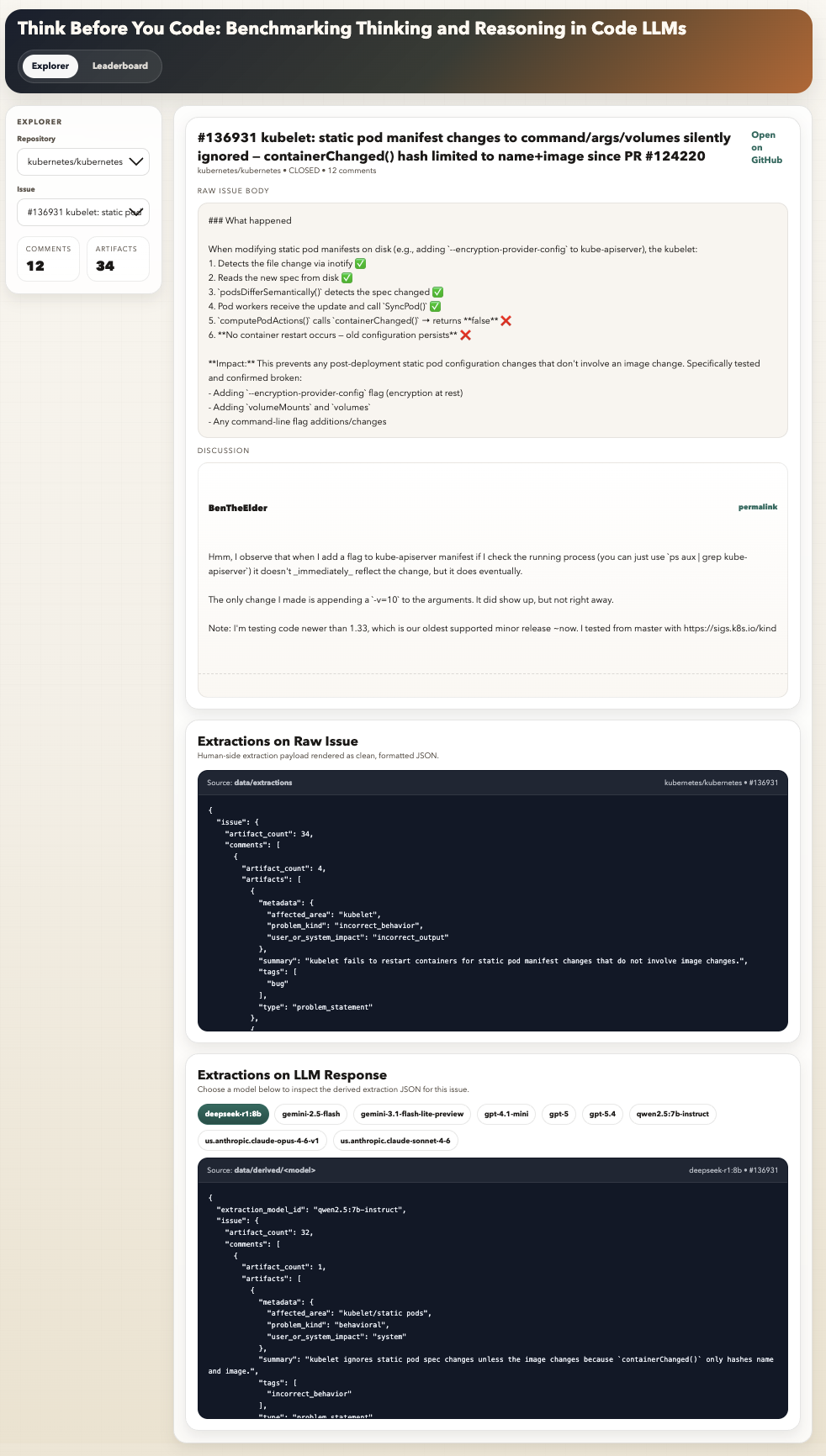}
    \caption{Interactive explorer for pulling up a repo and issue from the dataset. Users can view the issue, hummand comments, links to GitHub issues and the human and LLM artifacts across a wide range of models.}
    \label{fig:placeholder}
\end{figure}

\end{document}